\documentclass[fleqn]{llncs}
\usepackage{latexsym}
\usepackage{amssymb,amsmath}
\usepackage{stmaryrd}
\usepackage{graphicx}
\usepackage{hyperref}
\usepackage{phonetic}
\usepackage{xargs}
\usepackage[pdftex,dvipsnames]{xcolor}
\usepackage{listings}

\newcommand{\be}{\begin{enumerate}}
\newcommand{\ee}{\end{enumerate}}
\newcommand{\bi}{\begin{itemize}}
\newcommand{\ei}{\end{itemize}}
\newcommand{\bc}{\begin{center}}
\newcommand{\ec}{\end{center}}
\newcommand{\bsp}{\begin{sloppypar}}
\newcommand{\esp}{\end{sloppypar}}

\newtheorem{thm}{Theorem}[subsection]
\newtheorem{cor}[thm]{Corollary}

\newtheorem{rem}[thm]{Remark}

\newcommand{\sC}{\mbox{$\cal C$}}

\newcommand{\sE}{\mbox{$\cal E$}}

\newcommand{\sT}{\mbox{$\cal T$}}

\newcommand{\sV}{\mbox{$\cal V$}}

\renewcommand{\phi}{\varphi}

\newcommand{\churchqe}{$\mbox{\sc ctt}_{\rm qe}$}

\newcommand{\qzero}{${\cal Q}_0$}

\newcommand{\HOL}{$\mbox{\rm HOL}$}
\newcommand{\HL}{$\mbox{\rm HOL Light}$}
\newcommand{\HOLtwoP}{$\mbox{\rm HOL2P}$}
\newcommand{\HLQE}{$\mbox{\rm HOL Light QE}$}

\newcommand{\sembrack}[1]{\llbracket#1\rrbracket}
\newcommand{\synbrack}[1]{\ulcorner#1\urcorner}
\newcommand{\commabrack}[1]{\lfloor#1\rfloor}
\newcommand{\mname}[1]{\mbox{\sf #1}}

\newcommand{\mdot}{\mathrel.}
\newcommand{\tarrow}{\rightarrow}
\newcommand{\LambdaApp}{\lambda\,}
\newcommand{\Neg}{\neg}

\renewcommand{\And}{\wedge}
\newcommand{\Implies}{\supset}
\newcommand{\Or}{\vee}

\newcommand{\ForallApp}{\forall\,}

\newcommand{\ForsomeApp}{\exists\,}

\usepackage[colorinlistoftodos,textsize=tiny]{todonotes}
\newcommandx{\unsure}[2][1=]{\todo[linecolor=red,backgroundcolor=red!25,bordercolor=red,#1]{#2}}
\newcommandx{\change}[2][1=]{\todo[linecolor=blue,backgroundcolor=blue!25,bordercolor=blue,#1]{#2}}
\newcommandx{\info}[2][1=]{\todo[linecolor=OliveGreen,backgroundcolor=OliveGreen!25,bordercolor=OliveGreen,#1]{#2}}
\newcommandx{\improvement}[2][1=]{\todo[linecolor=Plum,backgroundcolor=Plum!25,bordercolor=Plum,#1]{#2}}

\title{{\HLQE}\thanks{To appear in the proceedings of the 9th
    International Conference on Interactive Theorem Proving (ITP
    2018).  This research was supported by NSERC.}}

\author{Jacques Carette, William M. Farmer, and Patrick Laskowski}

\institute{%
Computing and Software, McMaster University, Canada\\
\url{http://www.cas.mcmaster.ca/~carette}\\
\url{http://imps.mcmaster.ca/wmfarmer}\\[1.5ex]
12 May 2018
}

\begin{document}

\maketitle

\begin{abstract}
We are interested in algorithms that manipulate mathematical
expressions in mathematically meaningful ways. Expressions are
syntactic, but most logics do not allow one to discuss syntax.
{\churchqe} is a version of Church's type theory that includes
quotation and evaluation operators, akin to quote and eval
in the Lisp programming language.  Since the {\HOL} logic is also a
version of Church's type theory, we decided to add quotation and
evaluation to {\HL} to demonstrate the implementability of {\churchqe}
and the benefits of having quotation and evaluation in a proof
assistant.  The resulting system is called {\HLQE}.  Here we document
the design of {\HLQE} and the challenges that needed to be overcome.
The resulting implementation is freely available.
\end{abstract}

\iffalse 


\fi

\section{Introduction}\label{sec:introduction}

A \emph{syntax-based mathematical algorithm (SBMA)} manipulates
mathematical expressions in a meaningful way.  SBMAs
are commonplace in mathematics.  Examples include algorithms that
compute arithmetic operations by manipulating numerals, linear
transformations by manipulating matrices, and derivatives by
manipulating functional expressions.  Reasoning about the mathematical
meaning of an SBMA requires reasoning about the relationship between
how the expressions are manipulated by the SBMA and what the
manipulations mean.

We argue in~\cite{Farmer13} that the combination of quotation and
evaluation, along with appropriate inference rules, provides the means
to reason about the interplay between syntax and semantics, which is
what is needed for reasoning about SBMAs.  \emph{Quotation} is an
operation that maps an expression $e$ to a special value called a
\emph{syntactic value} that represents the syntax tree of $e$.
Quotation enables expressions to be manipulated as syntactic entities.
\emph{Evaluation} is an operation that maps a syntactic value $s$ to
the value of the expression that is represented by $s$.  Evaluation
enables meta-level reasoning via syntactic values to be reflected into
object-level reasoning.  Quotation and evaluation thus form an
infrastructure for integrating meta-level and object-level
reasoning. Quotation gives a form of \emph{reification} of
object-level values which allows introspection.  Along with inference
rules, this gives a certain amount of {\emph{logical reflection}};
evaluation adds to this some aspects of {\emph{computational
    reflection}}~\cite{Costantini02,Harrison95}.

Incorporating quotation and evaluation operators --- like quote and
eval in the Lisp programming language --- into a traditional logic
like first-order logic or simple type theory is not a straightforward
task.  Several challenging design problems stand in the way.  The
three design problems that most concern us are the following.  We will
write the quotation and evaluation operators applied to an expression
$e$ as $\synbrack{e}$ and $\sembrack{e}$, respectively.

\be

  \item \emph{Evaluation Problem.}  An evaluation operator is
    applicable to syntactic values that represent formulas and thus is
    effectively a truth predicate.  Hence, by the proof of Tarski's
    theorem on the undefinability of truth~\cite{Tarski35a}, if the
    evaluation operator is total in the context of a sufficiently
    strong theory (like first-order Peano arithmetic), then it is
    possible to express the liar paradox.  Therefore, the evaluation
    operator must be partial and the law of disquotation cannot hold
    universally (i.e., for some expressions $e$,
    $\sembrack{\synbrack{e}} \not= e$).  As a result, reasoning with
    evaluation can be cumbersome and leads to undefined expressions.

  \item \emph{Variable Problem.}  The variable $x$ is not free in the
    expression $\synbrack{x + 3}$ (or in any quotation).  However, $x$
    is free in $\sembrack{\synbrack{x + 3}}$ because
    $\sembrack{\synbrack{x + 3}} = x + 3$.  If the value of a constant
    $c$ is $\synbrack{x + 3}$, then $x$ is free in $\sembrack{c}$
    because $\sembrack{c} = \sembrack{\synbrack{x + 3}} = x + 3$.
    Hence, in the presence of an evaluation operator, whether or not a
    variable is free in an expression may depend on the values of the
    expression's components.  As a consequence, the substitution of an
    expression for the free occurrences of a variable in another
    expression depends on the semantics (as well as the syntax) of the
    expressions involved and must be integrated with the proof system
    for the logic.  That is, a logic with quotation and evaluation
    requires a semantics-dependent form of substitution in which side
    conditions, like whether a variable is free in an expression, are
    proved within the proof system.  This is a major departure from
    traditional logic.

  \item \emph{Double Substitution Problem.}  By the semantics of
    evaluation, the value of $\sembrack{e}$ is the \emph{value} of the
    expression whose syntax tree is represented by the \emph{value} of
    $e$.  Hence the semantics of evaluation involves a double
    valuation.  This is most apparent when the value of a variable
    involves a syntax tree that refers to the name of that same
    variable. For example, if the value of a variable $x$ is
    $\synbrack{x}$, then $\sembrack{x} = \sembrack{\synbrack{x}} = x =
    \synbrack{x}$.  Hence the substitution of $\synbrack{x}$ for $x$
    in $\sembrack{x}$ requires one substitution inside the argument of
    the evaluation operator and another substitution after the
    evaluation operator is eliminated.  This double substitution is
    another major departure from traditional logic.

\ee

{\churchqe}~\cite{Farmer16,Farmer18} is version of Church's type
theory~\cite{Church40} with quotation and evaluation that solves these
three design problems.  It is based on {\qzero}~\cite{Andrews02},
Peter Andrews' version of Church's type theory.  We believe
{\churchqe} is the first readily implementable version of simple type
theory that includes \emph{global} quotation and evaluation operators.  We
show in~\cite{Farmer18} that it is suitable for defining,
applying, and reasoning about SBMAs.

To demonstrate that {\churchqe} is indeed implementable, we have done
so by modifying \HL~\cite{Harrison09}, a compact implementation of the
HOL proof assistant~\cite{GordonMelham93}.  The resulting version of
{\HL} is called \HLQE.  Here we present its design, implementation,
and the challenges encountered.  ({\HOLtwoP}~\cite{Voelker07} is another
example of a logical system built by modifying {\HL}.)

The rest of the paper is organized as follows.
Section~\ref{sec:ctt-qe} presents the key ideas underlying {\churchqe}
and explains how {\churchqe} solves the three design problems.
Section~\ref{sec:hol-light} offers a brief overview of HOL
Light.  The \HLQE{} implementation is described in
section~\ref{sec:implementation}, and examples of how quotation and
evaluation are used in it are discussed in section~\ref{sec:examples}.
Section~\ref{sec:related-work} is devoted to related work.  And the
paper ends with some final remarks including a brief discussion on future work.

The major contributions of the work presented here are:

\be

  \item We show that the logical machinery for quotation and
    evaluation embodied in {\churchqe} can be straightforwardly
    implemented by modifying {\HL}.

  \item We produce an {\HOL}-style proof assistant with a built-in
    global reflection infrastructure for defining, applying, and
    proving properties about SBMAs.

  \item We demonstrate how this reflection infrastructure can be used
    to express formula schemas, such as the induction schema for
    first-order Peano arithmetic, as single formulas.

\ee

\section{${\rm CTT_{qe}}$}\label{sec:ctt-qe}

The syntax, semantics, and proof system of {\churchqe}
are defined in~\cite{Farmer18}.  Here
we will only introduce the definitions and results of that
are key to understanding how \HLQE{} implements {\churchqe}.  The
reader is encouraged to consult~\cite{Farmer18} when additional
details are required.

\subsection{Syntax}

{\churchqe} has the same machinery as {\qzero} plus an
inductive type $\epsilon$ of syntactic values, a partial quotation
operator, and a typed evaluation operator.

A \emph{type} of {\churchqe} is defined inductively by the following
formation rules:
%
%\vspace*{-1.5mm}
\be

  \item \emph{Type of individuals}: $\iota$ is a type.

  \item \emph{Type of truth values}: $\omicron$ is a type.

  \item \emph{Type of constructions}: $\epsilon$ is a type.

  \item \emph{Function type}: If $\alpha$ and $\beta$ are types, then
    $(\alpha \tarrow \beta)$ is a type.

\ee

\noindent
Let $\sT$ denote the set of types of {\churchqe}.  
A \emph{typed symbol} is a symbol with a subscript from $\sT$.  Let
$\sV$ be a set of typed symbols such that, for each $\alpha \in \sT$,
$\sV$ contains denumerably many typed symbols with subscript~$\alpha$.
A \emph{variable of type $\alpha$} of {\churchqe} is a member of $\sV$
with subscript~$\alpha$.  $\textbf{x}_\alpha, \textbf{y}_\alpha,
\textbf{z}_\alpha, \ldots$ are syntactic variables ranging over
variables of type $\alpha$. Let $\sC$ be a set of typed symbols
disjoint from $\sV$.  A \emph{constant of type $\alpha$} of
{\churchqe} is a member of $\sC$ with subscript~$\alpha$.
$\textbf{c}_\alpha, \textbf{d}_\alpha, \ldots$ are syntactic variables
ranging over constants of type~$\alpha$.  $\sC$ contains a set of
\emph{logical constants} that include $\mname{app}_{\epsilon \tarrow
  \epsilon \tarrow \epsilon}$, $\mname{abs}_{\epsilon \tarrow \epsilon
  \tarrow \epsilon}$, and $\mname{quo}_{\epsilon \tarrow \epsilon}$.

An \emph{expression of type $\alpha$} of {\churchqe} is defined
inductively by the formation rules below.  $\textbf{A}_\alpha,
\textbf{B}_\alpha, \textbf{C}_\alpha, \ldots$ are syntactic variables
ranging over expressions of type $\alpha$.  An expression is
\emph{eval-free} if it is constructed using just the first five
rules.
%
%\vspace*{-3mm}
\be

  \item \emph{Variable}: $\textbf{x}_\alpha$ is an expression of type
    $\alpha$.

  \item \emph{Constant}: $\textbf{c}_\alpha$ is an expression of type
    $\alpha$.

  \item \emph{Function application}: $(\textbf{F}_{\alpha \tarrow
    \beta} \, \textbf{A}_\alpha)$ is an expression of type $\beta$.

  \item \emph{Function abstraction}: $(\LambdaApp \textbf{x}_\alpha
    \mdot \textbf{B}_\beta)$ is an expression of type $\alpha \tarrow
    \beta$.

  \item \emph{Quotation}: $\synbrack{\textbf{A}_\alpha}$ is an
    expression of type $\epsilon$ if $\textbf{A}_\alpha$ is eval-free.

  \item \emph{Evaluation}: $\sembrack{\textbf{A}_\epsilon}_{{\bf
      B}_\beta}$ is an expression of type $\beta$.

\ee 

\noindent
The sole purpose of the second component $\textbf{B}_\beta$ in an
evaluation $\sembrack{\textbf{A}_\epsilon}_{{\bf B}_\beta}$ is to
establish the type of the evaluation; we will thus write
$\sembrack{\textbf{A}_\epsilon}_{{\bf B}_\beta}$ as
$\sembrack{\textbf{A}_\epsilon}_\beta$.

A \emph{construction} of {\churchqe} is an expression of type
$\epsilon$ defined inductively by:

\be

  \item $\synbrack{\textbf{x}_\alpha}$ is a construction.

  \item $\synbrack{\textbf{c}_\alpha}$ is a construction.

  \item If $\textbf{A}_\epsilon$ and $\textbf{B}_\epsilon$ are
    constructions, then $\mname{app}_{\epsilon \tarrow \epsilon
      \tarrow \epsilon} \, \textbf{A}_\epsilon \,
    \textbf{B}_\epsilon$, $\mname{abs}_{\epsilon \tarrow \epsilon
      \tarrow \epsilon} \, \textbf{A}_\epsilon \,
    \textbf{B}_\epsilon$, and $\mname{quo}_{\epsilon \tarrow \epsilon}
    \, \textbf{A}_\epsilon$ are constructions.

\ee

\noindent
The set of constructions is thus an inductive type whose base elements
are quotations of variables and constants, and whose constructors are
$\mname{app}_{\epsilon \tarrow \epsilon \tarrow \epsilon}$,
$\mname{abs}_{\epsilon \tarrow \epsilon \tarrow \epsilon}$, and
$\mname{quo}_{\epsilon \tarrow \epsilon}$.  As we will see shortly,
constructions serve as syntactic values.

Let $\sE$ be the function mapping eval-free expressions to
constructions that is defined inductively as follows:

\be

  \item $\sE(\textbf{x}_\alpha) = \synbrack{\textbf{x}_\alpha}$.

  \item $\sE(\textbf{c}_\alpha) = \synbrack{\textbf{c}_\alpha}$.

  \item $\sE(\textbf{F}_{\alpha \tarrow \beta} \, \textbf{A}_\alpha) =
    \mname{app}_{\epsilon \tarrow \epsilon \tarrow \epsilon} \,
    \sE(\textbf{F}_{\alpha \tarrow \beta}) \, \sE(\textbf{A}_\alpha)$.

  \item $\sE(\LambdaApp \textbf{x}_\alpha \mdot \textbf{B}_\beta) =
    \mname{abs}_{\epsilon \tarrow \epsilon \tarrow \epsilon} \,
    \sE(\textbf{x}_\alpha) \, \sE(\textbf{B}_\beta)$.

  \item $\sE(\synbrack{\textbf{A}_\alpha}) = \mname{quo}_{\epsilon
    \tarrow \epsilon} \, \sE(\textbf{A}_\alpha)$.

\ee

\noindent
When $\textbf{A}_\alpha$ is eval-free, $\sE(\textbf{A}_\alpha)$ is the
unique construction that represents the syntax tree of
$\textbf{A}_\alpha$.  That is, $\sE(\textbf{A}_\alpha)$ is a syntactic
value that represents how $\textbf{A}_\alpha$ is syntactically
constructed.  For every eval-free expression, there is a construction
that represents its syntax tree, but not every construction represents
the syntax tree of an eval-free expression.  For example,
$\mname{app}_{\epsilon \tarrow \epsilon \tarrow \epsilon} \,
\synbrack{\textbf{x}_\alpha} \, \synbrack{\textbf{x}_\alpha}$
represents the syntax tree of $(\textbf{x}_\alpha \,
\textbf{x}_\alpha)$ which is not an expression of {\churchqe} since
the types are mismatched.  A construction is \emph{proper} if it is in
the range of $\sE$, i.e., it represents the syntax tree of an
eval-free expression.

The purpose of $\sE$ is to define the semantics of quotation: the
meaning of $\synbrack{\textbf{A}_\alpha}$ is $\sE(\textbf{A}_\alpha)$.

\subsection{Semantics}

The semantics of {\churchqe} is based on Henkin-style general
models~\cite{Henkin50}.  An expression $\textbf{A}_\epsilon$ of type
$\epsilon$ denotes a construction, and when $\textbf{A}_\epsilon$ is a
construction, it denotes itself.  The semantics of the quotation and
evaluation operators are defined so that the following two theorems
hold:

\begin{thm}[Law of Quotation] \label{thm:sem-quotation}
$\synbrack{\textbf{A}_\alpha} = \sE(\textbf{A}_\alpha)$ is valid in
  {\churchqe}.
\end{thm}

\begin{cor}$\synbrack{\textbf{A}_\alpha} = \synbrack{\textbf{B}_\alpha}$ iff $\textbf{A}_\alpha$ and $\textbf{B}_\alpha$ are identical expressions.
\end{cor}

\begin{thm}[Law of Disquotation] \label{thm:sem-disquotation}
$\sembrack{\synbrack{\textbf{A}_\alpha}}_\alpha = \textbf{A}_\alpha$
  is valid in {\churchqe}.
\end{thm}

\begin{rem}\em
Notice that this is not the full Law of Disquotation, since only
eval-free expressions can be quoted.  As a result of this restriction,
the liar paradox is not expressible in {\churchqe} and the Evaluation
Problem mentioned above is effectively solved.
\end{rem}

\subsection{Quasiquotation}

Quasiquotation is a parameterized form of quotation in which the
parameters serve as holes in a quotation that are filled with
expressions that denote syntactic values.  It is a very powerful
syntactic device for specifying expressions and defining macros.
Quasiquotation was introduced by Willard Van Orman Quine in 1940 in
the first version of his book \emph{Mathematical
  Logic}~\cite{Quine03}.  It has been extensively employed in the Lisp
family of programming languages~\cite{Bawden99}\footnote{In Lisp, the
  standard symbol for quasiquotation is the backquote ({\tt `})
  symbol, and thus in Lisp, quasiquotation is usually called
  \emph{backquote}.}, and from there to other families of
programming languages, most notably the ML family.

In {\churchqe}, constructing a large quotation from smaller quotations
can be tedious because it requires many applications of the syntax
constructors $\mname{app}_{\epsilon \tarrow \epsilon \tarrow
  \epsilon}$, $\mname{abs}_{\epsilon \tarrow \epsilon \tarrow
  \epsilon}$, and $\mname{quo}_{\epsilon \tarrow \epsilon}$.
Quasiquotation alleviates this problem.
It can be defined straightforwardly in
{\churchqe}.  However, quasiquotation is not part of the official
syntax of {\churchqe}; it is just a notational device used to write
{\churchqe} expressions in a compact form.

As an example, consider $\synbrack{\Neg(\textbf{A}_o \And
  \commabrack{\textbf{B}_\epsilon})}$. Here $\commabrack{\textbf{B}_\epsilon}$
is a \emph{hole} or \emph{antiquotation}. Assume that
$\textbf{A}_o$ contains no holes.  $\synbrack{\Neg(\textbf{A}_o \And
  \commabrack{\textbf{B}_\epsilon})}$ is then an abbreviation for the
verbose expression
\[\mname{app}_{\epsilon \tarrow
  \epsilon \tarrow \epsilon} \, \synbrack{\Neg_{o \tarrow o}} \,
(\mname{app}_{\epsilon \tarrow \epsilon \tarrow \epsilon} \,
(\mname{app}_{\epsilon \tarrow \epsilon \tarrow \epsilon}
\synbrack{\wedge_{o \tarrow o \tarrow o}} \, \synbrack{\textbf{A}_o})
\, \textbf{B}_\epsilon).\] $\synbrack{\Neg(\textbf{A}_o \And
  \commabrack{\textbf{B}_\epsilon})}$ represents the the syntax tree
of a negated conjunction in which the part of the tree corresponding
to the second conjunct is replaced by the syntax tree represented by
$\textbf{B}_\epsilon$.  If $\textbf{B}_\epsilon$ is a quotation
$\synbrack{\textbf{C}_o}$, then the quasiquotation
$\synbrack{\Neg(\textbf{A}_o \And
  \commabrack{\synbrack{\textbf{C}_o}})}$ is \emph{equivalent} to the
quotation $\synbrack{\Neg(\textbf{A}_o \And \textbf{C}_o)}$.

\subsection{Proof System}\label{subsec:cttqe-proof-system}

The proof system for {\churchqe} consists of the axioms for {\qzero},
the single rule of inference for {\qzero}, and additional
axioms~\cite[B1--B13]{Farmer18} that define the logical constants
of {\churchqe} (B1--B4, B5, B7), specify $\epsilon$ as an inductive type
(B4, B6), state the properties of quotation and evaluation (B8, B10),
and extend the rules for beta-reduction (B9, B11--13).  We prove
in~\cite{Farmer18} that this proof system is sound for all
formulas and complete for eval-free formulas.

The axioms that express the properties of quotation and evaluation are:

\medskip

\noindent
\begin{minipage}{\textwidth}
  \noindent\textbf{B8 (Properties of Quotation)}
  \be

    \item $\synbrack{\textbf{F}_{\alpha \tarrow \beta} \,
      \textbf{A}_\alpha} = \mname{app}_{\epsilon \tarrow \epsilon
      \tarrow \epsilon} \, \synbrack{\textbf{F}_{\alpha \tarrow
        \beta}} \, \synbrack{\textbf{A}_\alpha}$.

    \item $\synbrack{\LambdaApp \textbf{x}_\alpha \mdot
      \textbf{B}_\beta} = \mname{abs}_{\epsilon \tarrow \epsilon
      \tarrow \epsilon} \, \synbrack{\textbf{x}_\alpha} \,
      \synbrack{\textbf{B}_\beta}$.

    \item $\synbrack{\synbrack{\textbf{A}_\alpha}} =
      \mname{quo}_{\epsilon \tarrow \epsilon} \,
      \synbrack{\textbf{A}_\alpha}$.

  \ee

  \noindent\textbf{B10 (Properties of Evaluation)}

  \be

    \item $\sembrack{\synbrack{\textbf{x}_\alpha}}_\alpha = \textbf{x}_\alpha$.

    \item $\sembrack{\synbrack{\textbf{c}_\alpha}}_\alpha = \textbf{c}_\alpha$.

    \item $(\mname{is-expr}_{\epsilon \tarrow o}^{\alpha \tarrow \beta}
      \, \textbf{A}_\epsilon \And \mname{is-expr}_{\epsilon \tarrow o}^{\alpha}
      \, \textbf{B}_\epsilon) \Implies
          \sembrack{\mname{app}_{\epsilon \tarrow \epsilon
          \tarrow \epsilon} \, \textbf{A}_\epsilon \,
        \textbf{B}_\epsilon}_{\beta} =
      \sembrack{\textbf{A}_\epsilon}_{\alpha \tarrow \beta} \,
      \sembrack{\textbf{B}_\epsilon}_{\alpha}$.

    \item $(\mname{is-expr}_{\epsilon \tarrow o}^{\beta} \,
      \textbf{A}_\epsilon \And \Neg(\mname{is-free-in}_{\epsilon \tarrow
        \epsilon \tarrow o} \, \synbrack{\textbf{x}_\alpha} \,
      \synbrack{\textbf{A}_\epsilon})) \Implies\\[.5ex]
      \hspace*{2ex} \sembrack{\mname{abs}_{\epsilon \tarrow \epsilon
          \tarrow \epsilon} \, \synbrack{\textbf{x}_\alpha} \,
        \textbf{A}_\epsilon}_{\alpha \tarrow \beta} = \LambdaApp
      \textbf{x}_\alpha \mdot \sembrack{\textbf{A}_\epsilon}_\beta$.

    \item $\mname{is-expr}_{\epsilon \tarrow o}^\epsilon \,
      \textbf{A}_\epsilon \Implies \sembrack{\mname{quo}_{\epsilon
          \tarrow \epsilon} \, \textbf{A}_\epsilon}_\epsilon =
      \textbf{A}_\epsilon$.

  \ee

\end{minipage}

\medskip

\noindent
The axioms for extending the rules  for beta-reduction are:

\medskip

\noindent
\begin{minipage}{\textwidth}
  \noindent\textbf{B9 (Beta-Reduction for Quotations)}

  \be

    \item[] $(\LambdaApp \textbf{x}_\alpha \mdot
      \synbrack{\textbf{B}_\beta}) \, \textbf{A}_\alpha) =
      \synbrack{\textbf{B}_\beta}$.

  \ee

  \noindent\textbf{B11 (Beta-Reduction for Evaluations)}
  \be

    \item $(\LambdaApp \textbf{x}_\alpha \mdot
      \sembrack{\textbf{B}_\epsilon}_\beta) \, \textbf{x}_\alpha =
      \sembrack{\textbf{B}_\epsilon}_\beta$.
 
    \item $(\mname{is-expr}_{\epsilon \tarrow o}^{\beta} \,
      ((\LambdaApp \textbf{x}_\alpha \mdot \textbf{B}_\epsilon) \,
      \textbf{A}_\alpha) \And %\\[.5ex] \hspace*{0.5ex}
      \Neg(\mname{is-free-in}_{\epsilon \tarrow \epsilon \tarrow o}
      \, \synbrack{\textbf{x}_\alpha} \, ((\LambdaApp
      \textbf{x}_\alpha \mdot \textbf{B}_\epsilon) \,
      \textbf{A}_\alpha))) \Implies \\[.5ex]
      \hspace*{2ex} (\LambdaApp \textbf{x}_\alpha \mdot
      \sembrack{\textbf{B}_\epsilon}_\beta) \, \textbf{A}_\alpha =
      \sembrack{(\LambdaApp \textbf{x}_\alpha \mdot
        \textbf{B}_\epsilon) \, \textbf{A}_\alpha}_\beta$.

  \ee

  \noindent\textbf{B12 (``Not Free In'' means ``Not Effective In'')}

  \bi

    \item[] $\Neg\mname{IS-EFFECTIVE-IN}(\textbf{x}_\alpha,\textbf{B}_\beta)$\\
      where $\textbf{B}_\beta$ is eval-free and
      $\textbf{x}_\alpha$ is not free in $\textbf{B}_\beta$.

  \ei

  \noindent\textbf{B13 (Beta-Reduction for Function Abstractions)}

  \bi

    \item[] $(\Neg \mname{IS-EFFECTIVE-IN}(\textbf{y}_\beta,\textbf{A}_\alpha)\Or 
      \Neg \mname{IS-EFFECTIVE-IN}(\textbf{x}_\alpha,\textbf{B}_\gamma)) \Implies {}\\ 
      \hspace*{2ex}(\LambdaApp \textbf{x}_\alpha \mdot 
      \LambdaApp \textbf{y}_\beta \mdot \textbf{B}_\gamma) \, \textbf{A}_\alpha =
      \LambdaApp \textbf{y}_\beta \mdot 
      ((\LambdaApp \textbf{x}_\alpha \mdot \textbf{B}_\gamma) \, \textbf{A}_\alpha)$\\
      where $\textbf{x}_\alpha$ and $\textbf{y}_\beta$ are distinct.

   \ei

\end{minipage}

\medskip

Substitution is performed using
the properties of beta-reduction as Andrews does in the proof system
for {\qzero}~\cite[p.~213]{Andrews02}.  The following three
beta-reduction cases require discussion:

%\vspace*{-2.5mm}
\be

  \item $(\LambdaApp \textbf{x}_\alpha \mdot \LambdaApp
    \textbf{y}_\beta \mdot \textbf{B}_\gamma) \, \textbf{A}_\alpha$
    where $\textbf{x}_\alpha$ and $\textbf{y}_\beta$ are distinct.

  \item $(\LambdaApp \textbf{x}_\alpha \mdot
      \synbrack{\textbf{B}_\beta}) \, \textbf{A}_\alpha$.

  \item $(\LambdaApp \textbf{x}_\alpha \mdot
    \sembrack{\textbf{B}_\epsilon}_\beta) \, \textbf{A}_\alpha$.

\ee

The first case can normally be reduced when either (1)
$\textbf{y}_\beta$ is not free in $\textbf{A}_\alpha$ or (2)
$\textbf{x}_\alpha$ is not free in $\textbf{B}_\gamma$.  However, due
to the Variable Problem mentioned before, it is only possible to
syntactically check whether a ``variable is not free in an
expression'' when the expression is eval-free.  Our solution
is to replace the syntactic notion of ``a variable is free in
an expression'' by the semantic notion of ``a variable is effective in
an expression'' when the expression is not necessarily eval-free, and
use Axiom B13 to perform the beta-reduction.  

``$\textbf{x}_\alpha$ is effective in $\textbf{B}_\beta$'' means the
value of $\textbf{B}_\beta$ depends on the value of
$\textbf{x}_\alpha$.  Clearly, if $\textbf{B}_\beta$ is eval-free,
``$\textbf{x}_\alpha$ is effective in $\textbf{B}_\beta$'' implies
``$\textbf{x}_\alpha$ is free in $\textbf{B}_\beta$''.  However,
``$\textbf{x}_\alpha$ is effective in ${B}_\beta$'' is a refinement of
``$\textbf{x}_\alpha$ is free in $\textbf{B}_\beta$'' on eval-free
expressions since $\textbf{x}_\alpha$ is free in $\textbf{x}_\alpha =
\textbf{x}_\alpha$, but $\textbf{x}_\alpha$ is not effective in
$\textbf{x}_\alpha = \textbf{x}_\alpha$.  ``$\textbf{x}_\alpha$ is
effective in $\textbf{B}_\beta$'' is expressed in {\churchqe} as
$\mname{IS-EFFECTIVE-IN}(\textbf{x}_\alpha,\textbf{B}_\beta)$, an
abbreviation for \[\ForsomeApp \textbf{y}_\alpha \mdot ((\LambdaApp
\textbf{x}_\alpha \mdot \textbf{B}_\beta) \, \textbf{y}_\alpha \not=
\textbf{B}_\beta)\] where $\textbf{y}_\alpha$ is any variable of type
$\alpha$ that differs from $\textbf{x}_\alpha$.

The second case is simple since a quotation cannot be modified by
substitution --- it is effectively the same as a constant. Thus
beta-reduction is performed without changing
$\synbrack{\textbf{B}_\beta}$ as shown in Axiom B9 above.

The third case is handled by Axioms B11.1 and B11.2.  B11.1 deals with
the trivial case when $\textbf{A}_\alpha$ is the bound variable
$\textbf{x}_\alpha$ itself.  B11.2 deals with the other much more
complicated situation.  The condition
\[\Neg(\mname{is-free-in}_{\epsilon \tarrow \epsilon \tarrow o}
      \, \synbrack{\textbf{x}_\alpha} \, ((\LambdaApp
      \textbf{x}_\alpha \mdot \textbf{B}_\epsilon) \,
      \textbf{A}_\alpha))\] guarantees that there is no
      \emph{double substitution}.  $\mname{is-free-in}_{\epsilon \tarrow
        \epsilon \tarrow o}$ is a logical constant of {\churchqe} such
      that $\mname{is-free-in}_{\epsilon \tarrow \epsilon \tarrow o}
      \, \synbrack{\textbf{x}_\alpha} \,
      \synbrack{\textbf{B}_\beta}$ says that the variable
      $\textbf{x}_\alpha$ is free in the (eval-free) expression
      $\textbf{B}_\beta$.

Thus we see that substitution in {\churchqe} in the presence of
evaluations may require proving semantic side conditions of the
following two forms:

\be

  \item $\Neg
    \mname{IS-EFFECTIVE-IN}(\textbf{x}_\alpha,\textbf{B}_\beta)$.

  \item $\mname{is-free-in}_{\epsilon \tarrow \epsilon \tarrow o} \,
    \synbrack{\textbf{x}_\alpha} \, \synbrack{\textbf{B}_\beta}$.

\ee

\subsection{The Three Design Problems} 

To recap,
{\churchqe} solves the three design problems given in
section~\ref{sec:introduction}.  The Evaluation Problem is
avoided by restricting the quotation operator to eval-free expressions
and thus making it impossible to express the liar paradox.  The
Variable Problem is overcome by modifying Andrews' beta-reduction
axioms.  The Double Substitution Problem is eluded
by using a beta-reduction axiom for evaluations that excludes
beta-reductions that embody a double substitution.

\section{HOL Light}\label{sec:hol-light}

\HL~\cite{Harrison09} is an open-source proof assistant developed by
John Harrison.  It implements a logic (HOL) which is a version of
Church's type theory.  It is a simple implementation of the HOL proof
assistant~\cite{GordonMelham93} written in OCaml and hosted on GitHub
at \url{https://github.com/jrh13/hol-light/}.  Although it is a
relatively small system, it has been used to formalize many kinds of
mathematics and to check many proofs including the lion's share of Tom
Hales' proof of the Kepler conjecture~\cite{HalesEtAl17}.

{\HL} is very well suited to serve as a foundation on which to build
an implementation of {\churchqe}: First, it is an open-source system
that can be freely modified as long as certain very minimal conditions
are satisfied.  Second, it is an implementation of a version of simple
type theory that is essentially {\qzero}, the version of Church's type
theory underlying {\churchqe}, plus (1) polymorphic type variables,
(2) an axiom of choice expressed by asserting that the Hilbert
$\epsilon$ operator is a choice (indefinite description) operator, and
(3) an axiom of infinity that asserts that \texttt{ind}, the type of
individuals, is infinite~\cite{Harrison09}.  The type variables in the
implemented logic are not a hindrance; they actually facilitate the
implementation of {\churchqe}.  The presence of the axioms of choice
and infinity in {\HL} alter the semantics of {\churchqe} without
compromising in any way the semantics of quotation and evaluation.
And third, \HL{} supports the definition of inductive types so that
$\epsilon$ can be straightforwardly defined.

\section{Implementation}\label{sec:implementation}

\subsection{Overview}

{\HLQE} was implemented in four stages:

\be

  \item The set of terms was extended so that {\churchqe}
    expressions could be mapped to {\HL} terms.  This required the
    introduction of \texttt{epsilon}, the type of constructions, and
    term constructors for quotations and evaluations.  See
    subsection~\ref{subsec:mapping}.

  \item The proof system was modified to include the machinery
    in {\churchqe} for reasoning about quotations and evaluations.
    This required adding new rules of inference and modifying
    the \texttt{INST} rule of inference that simultaneously
    substitutes terms $t_1,\ldots,t_n$ for the free variables
    $x_1,\ldots,x_n$ in a sequent.  See
    subsection~\ref{subsec:hlqe-proof-system}.

  \item Machinery --- consisting of {\HOL} function definitions,
    tactics, and theorems --- was created for supporting reasoning
    about quotations and evaluations in the new system.  See
    subsection~\ref{subsec:machinery}.

  \item Examples were developed in the new system to test the
    implementation and to demonstrate the benefits of having quotation
    and evaluation in higher-order logic.  See section~\ref{sec:examples}.

\ee

\noindent
The first and second stages have been completed; both stages involved
modifying the kernel of {\HL}.  The third stage is sufficiently
complete to enable our examples in section~\ref{sec:examples} to work
well, and did not involve any further changes to the {\HL} kernel.  We
do expect that adding further examples, which is ongoing, will require
additional machinery but no changes to the kernel.

The {\HLQE} system was developed by the third author under the
supervision of the first two authors on an undergraduate NSERC USRA
research project at McMaster University and is available at
\[\mbox{\url{https://github.com/JacquesCarette/hol-light}}.\]
It should be further remarked that our fork, from late April 2017, is not fully
up-to-date with respect to {\HL}. In particular, this means that it is best to
compile it with \textsc{OCaml 4.03.0} and \textsc{camlp5 6.16},
both available from \textsc{opam}.

To run {\HLQE}, execute the following commands in {\HLQE} top-level
directory named \texttt{hol\_light}:

\begin{lstlisting}
1) install opam
2) opam init --comp 4.03.0                                                      
3) opam install "camlp5=6.16" 
5) opam `eval config env`
5) cd hol_light
6) make
7) run ocaml via                                                                
     ocaml -I `camlp5 -where` camlp5o.cma                                        
8) #use "hol.ml";;
   #use "Constructions/epsilon.ml";;
   #use "Constructions/pseudoquotation.ml";;
   #use "Constructions/QuotationTactics.ml";;
\end{lstlisting}
\noindent Each test can be run by an appropriate further
\lstinline|#use| statement.

\subsection{Mapping of ${\rm CTT_{qe}}$ Expressions to HOL Terms}\label{subsec:mapping}

\begin{table}[b]
\bc
\begin{tabular}{|lll|}
\hline
\textbf{${\bf CTT_{qe}}$ Type $\alpha$} \hspace*{2ex}
  & \textbf{HOL Type $\mu(\alpha)$}
  & \textbf{Abbreviation for $\mu(\alpha)$}\\
$\omicron$ & \texttt{Tyapp("bool",[])} & \texttt{bool}\\
$\iota$ & \texttt{Tyapp("ind",[])} & \texttt{ind}\\
$\epsilon$ & \texttt{Tyapp("epsilon",[])} & \texttt{epsilon}\\
$\beta \tarrow \gamma$ 
  & \texttt{Tyapp("fun",[\mbox{$\mu(\beta),\mu(\gamma)$}])} \hspace*{2ex}
  & \texttt{\mbox{$\mu(\beta)$}->\mbox{$\mu(\gamma)$}}\\  
\hline
\end{tabular}
\ec
\caption{Mapping of {\churchqe} Types to {\HOL} Types}\label{tab:types} 
\end{table}

\begin{table}[t]
\bc
\begin{tabular}{|ll|}
\hline
\textbf{${\bf CTT_{qe}}$ Expression $e$} \hspace*{2ex}
  & \textbf{HOL Term $\nu(e)$}\\
$\textbf{x}_\alpha$
  & \texttt{Var("x",\mbox{$\mu(\alpha)$})}\\
$\textbf{c}_\alpha$
  & \texttt{Const("c",\mbox{$\mu(\alpha)$})}\\
$\mname{=}_{\alpha \tarrow \alpha \tarrow o}$
  & \texttt{Const("=",\texttt{a\_ty\_var->a\_ty\_var->bool})}\\
$(\textbf{F}_{\alpha \tarrow \beta} \, \textbf{A}_\alpha)$
  & \texttt{Comb(\mbox{\rm $\nu(\textbf{F}_{\alpha \tarrow \beta}),\nu(\textbf{A}_\alpha)$})}\\
$(\LambdaApp \textbf{x}_\alpha \mdot \textbf{B}_\beta)$
  & \texttt{Abs(Var("x",\mbox{$\mu(\alpha)$}),\mbox{\rm $\nu(\textbf{B}_\beta)$})}\\
$\synbrack{\textbf{A}_\alpha}$
  & \texttt{Quote(\mbox{\rm $\nu(\textbf{A}_\alpha),\mu(\alpha)$})}\\
$\sembrack{\textbf{A}_\epsilon}_{{\bf B}_\beta}$
  & \texttt{Eval(\mbox{\rm $\nu(\textbf{A}_\epsilon),\mu(\beta)$})}\\
\hline
\end{tabular}
\ec
\caption{Mapping of {\churchqe} Expressions to {\HOL} Terms}\label{tab:expressions} 
\end{table}

Tables~\ref{tab:types} and~\ref{tab:expressions} illustrate how the
{\churchqe} types and expressions are mapped to the {\HOL} types and
terms, respectively.  The {\HOL} types and terms are written in the
the internal representation employed in {\HLQE}.  The type
\texttt{epsilon} and the term constructors \texttt{Quote} and
\texttt{Eval} are additions to {\HL} explained below.  Since
{\churchqe} does not have type variables, it has a logical constant
$\mname{=}_{\alpha \tarrow \alpha \tarrow o}$ representing equality
for each $\alpha \in \sT$.  The members of this family of constants
are all mapped to a single {\HOL} constant with the polymorphic type
\texttt{a\_ty\_var->a\_ty\_var->bool} where \texttt{a\_ty\_var} is any
chosen {\HOL} type variable.  

The other logical constants of {\churchqe}~\cite[Table 1]{Farmer18}
are not mapped to primitive {\HOL} constants.  $\mname{app}_{\epsilon
  \tarrow \epsilon \tarrow \epsilon}$, $\mname{abs}_{\epsilon \tarrow
  \epsilon \tarrow \epsilon}$, and $\mname{quo}_{\epsilon \tarrow
  \epsilon}$ are implemented by \texttt{App}, \texttt{Abs}, and
\texttt{Quo}, constructors for the inductive type \texttt{epsilon}
given below.  The remaining logical constants are predicates on
constructions that are implemented by {\HOL} functions.
The {\churchqe} type $\epsilon$ is the type of constructions, the
syntactic values that represent the syntax trees of eval-free
expressions.  $\epsilon$ is formalized as an inductive type
\texttt{epsilon}.  Since types are components of terms in
{\HL}, an inductive type \texttt{type} of syntax values for
{\HLQE} types (which are the same as {\HOL} types) is also
needed.  Specifically:

\begin{lstlisting}
define_type "type = TyVar string
                  | TyBase string
                  | TyMonoCons string type
                  | TyBiCons string type type"

define_type "epsilon = QuoVar string type 
                     | QuoConst string type
                     | App epsilon epsilon
                     | Abs epsilon epsilon
                     | Quo epsilon"
\end{lstlisting}

\noindent
Terms of type \texttt{type} denote the syntax trees of {\HLQE} types,
while the terms of type \texttt{epsilon} denote the syntax trees of
those terms that are eval-free.

The OCaml type of {\HOL} types in {\HLQE}

\begin{lstlisting}
type hol_type = Tyvar of string
              | Tyapp of string * hol_type list
\end{lstlisting}

\noindent
is the same as in {\HL}, but the OCaml type of {\HOL} terms in {\HLQE}

\begin{lstlisting}
type term = Var of string * hol_type
          | Const of string * hol_type
          | Comb of term * term
          | Abs of term * term
          | Quote of term * hol_type
          | Hole of term * hol_type
          | Eval of term * hol_type
\end{lstlisting}

\noindent
has three new constructors: \texttt{Quote}, \texttt{Hole}, and
\texttt{Eval}.

\texttt{Quote} constructs a quotation of type \texttt{epsilon} with
components $t$ and $\alpha$ from a term $t$ of type $\alpha$ that is
is eval-free.  \texttt{Eval} constructs an evaluation of type $\alpha$
with components $t$ and $\alpha$ from a term $t$ of type
\texttt{epsilon} and a type $\alpha$.  \texttt{Hole} is used to
construct ``holes'' of type \texttt{epsilon} in a quasiquotation as
described in~\cite{Farmer18}.  A quotation that contains
holes is a quasiquotation, while a quotation without any holes is a
normal quotation.  The construction of terms has been
modified to allow a hole (of type \texttt{epsilon}) to be used where a
term of some other type is expected.

The external representation of a quotation \texttt{Quote(t,ty)} is
\texttt{Q\_ t \_Q}.  Similarly, the external representation of a hole
\texttt{Hole(t,ty)} is \texttt{H\_ t \_H}.  The external
representation of an evaluation \texttt{Eval(t,ty)} is $\texttt{eval
  t to ty}.$

\subsection{Modification of the HOL Light Proof System}\label{subsec:hlqe-proof-system}

The proof system for {\churchqe} is obtained by extending {\qzero}'s
with additional axioms B1--B13 (see~\ref{subsec:cttqe-proof-system}).
Since {\qzero} and
{\HL} are both complete (with respect to the semantics of Henkin-style
general models), {\HL} includes the reasoning capabilities of the proof
system for {\qzero} but not the reasoning capabilities
embodied in the B1--B13 axioms, which must be
implemented in {\HLQE} as follows.
First, the logical constants defined by
Axioms B1--B4, B5, and B7 are defined in {\HLQE} as {\HOL} functions.
Second, the no junk (B6) and no confusion (B4) requirements for
$\epsilon$ are automatic consequences of defining \texttt{epsilon} as
an inductive type.  Third, Axiom B9 is implemented directly
in the {\HL} code for substitution.  Fourth, the remaining axioms, B8
and B10--B13 are implemented by new rules of inference in as
shown in Table~\ref{tab:axioms}.

%\vspace{-2mm}
\begin{table}
\bc
\begin{tabular}{|ll|}
\hline
${\bf CTT_{qe}}$ \textbf{Axioms}                & \textbf{NewRules of Inference}\\
B8 (Properties of Quotation)                     & \texttt{LAW\_OF\_QUO}\\
B10 (Properties of Evaluation)                   & \texttt{}\\
B10.1                                            & \texttt{VAR\_DISQUO}\\
B10.2                                            & \texttt{CONST\_DISQUO}\\
B10.3                                            & \texttt{APP\_SPLIT}\\
B10.4                                            & \texttt{ABS\_SPLIT}\\
B10.5                                            & \texttt{QUOTABLE}\\
B11 (Beta-Reduction for Evaluations)             & \texttt{}\\
B11.1                                            & \texttt{BETA\_EVAL}\\
B11.2                                            & \texttt{BETA\_REVAL}\\
B12 (``Not Free In'' means ``Not Effective In'') & \texttt{NOT\_FREE\_OR\_EFFECTIVE\_IN}\\
B13 (Beta-Reduction for Function Abstractions)  & \texttt{NEITHER\_EFFECTIVE}\\
\hline
\end{tabular}
\ec
\caption{New Inference Rules in {\HLQE}}\label{tab:axioms} 
\end{table}

%\vspace*{-7mm}
The \texttt{INST} rule of inference is also modified.  This rule
simultaneously substitutes a list of terms for a list of variables in
a sequent.  The substitution function \texttt{vsubst} defined in the
{\HL} kernel is modified so that it works like substitution (via
beta-reduction  rules) does in {\churchqe}.  The main changes are:

\be

  \item A substitution of a term \texttt{t} for a variable \texttt{x}
    in a function abstraction \texttt{Abs(y,s)} is performed as usual
    if (1) \texttt{t} is eval-free and \texttt{x} is not free in
    \texttt{t}, (2) there is a theorem that says \texttt{x} is not
    effective in \texttt{t}, (3) \texttt{s} is eval-free and
    \texttt{x} is not free in \texttt{s}, or (4) there is a theorem
    that says \texttt{x} is not effective in \texttt{s}.  Otherwise,
    if \texttt{s} or \texttt{t} is not eval-free, the substitution
    fails and if \texttt{s} and \texttt{t} are eval-free, the variable
    \texttt{x} is renamed and the substitution is continued.

  \item A substitution of a term \texttt{t} for a variable \texttt{x}
    in a quotation \texttt{Quote(e,ty)} where \texttt{e} does not
    contain any holes (i.e., terms of the form \texttt{Hole(e',ty')})
    returns \texttt{Quote(e,ty)} unchanged (as stated in Axiom B9).
    If \texttt{e} does contain holes, then \texttt{t} is substituted
    for the variable \texttt{x} in the holes in \texttt{Quote(e,ty)}.

  \item A substitution of a term \texttt{t} for a variable \texttt{x}
    in an evaluation \texttt{Eval(e,ty)} returns (1)
    \texttt{Eval(e,ty)} when \texttt{t} is \texttt{x} and (2) the
    function abstraction application
    \texttt{Comb(Abs(x,Eval(e,ty)),t)} otherwise.  (1) is valid by
    Axiom B11.1.  When (2) happens, this part of the substitution is
    finished and the user can possibly continue it by applying
    \texttt{BETA\_REVAL}, the rule of inference corresponding to Axiom
    B11.2.

\ee

\subsection{Creation of Support Machinery}\label{subsec:machinery}

The {\HLQE} system contains a number of {\HOL} functions, tactics, and
theorems that are useful for reasoning about constructions,
quotations, and evaluations.  An important example is the {\HOL}
function \texttt{isExprType} that implements the {\churchqe} family of
logical constants $\mname{is-expr}_{\epsilon \tarrow o}^{\alpha}$
where $\alpha$ ranges over members of $\sT$.  This function takes
terms $s_1$ and $s_1$ of type \texttt{epsilon} and \texttt{type},
respectively, and returns true iff $s_1$ represents the syntax tree of
a term $t$, $s_2$ represents the syntax tree of a type $\alpha$, and
$t$ is of type $\alpha$.

\subsection{Metatheorems}

We state three important metatheorems about {\HLQE}.  The proofs of
these metatheorems are straightforward but also tedious.  We label the
metatheorems as conjectures since their proofs have not yet been fully
written down.

\begin{conjecture}\bsp
Every formula provable in {\HL}'s proof system is also provable in
{\HLQE}'s proof system.\esp
\end{conjecture}

\noindent
\emph{Proof sketch.}  {\HLQE}'s proof system extends {\HL}'s proof
system with new machinery for reasoning about quotations and
evaluations. Thus every {\HL} proof remains valid in {\HLQE}. $\Box$

\bigskip

\noindent
Note: All the proofs loaded with the {\HL} system continue to be valid
when loaded in {\HLQE}.  A further test for the future would be to
load a variety of large {\HL} proofs in {\HLQE} to check that their
validity is preserved.

\begin{conjecture}
The proof system for {\HLQE} is sound for all formulas and complete
for all eval-free formulas.
\end{conjecture}

\noindent
\emph{Proof sketch.}  The analog of this statement for {\churchqe} is
proved in~\cite{Farmer18}.  It should be possible to prove this
conjecture by just imitating the proof for {\churchqe}. $\Box$

\begin{conjecture}\bsp
{\HLQE} is a model-theoretic conservative extension of {\HL}.\esp
\end{conjecture}

\noindent
\emph{Proof sketch.}  A model of {\HLQE} is a model of {\HL} with
definitions of the type $\epsilon$ and several constants and
interpretations for the (quasi)quotation and evaluation operators.
These additions do not impinge upon the semantics of {\HL}; hence
every model of {\HL} can be expanded to a model of the {\HLQE}, which
is the meaning of the conjecture. $\Box$

\section{Examples}\label{sec:examples}

We present two examples that illustrate its capabilities by expressing,
instantiating, and proving formula schemas in {\HLQE}.

\subsection{Law of Excluded Middle}

The \emph{law of excluded middle (LEM)} is expressed as the formula
schema $A \Or \Neg A$ where $A$ is a syntactic variable ranging over
all formulas.  Each instance of LEM is a theorem of {\HOL}, but LEM
cannot be expressed in {\HOL} as a single formula.  However, LEM can
be formalized in {\churchqe} as the universal statement
\[\ForallApp x_\epsilon \mdot 
\mname{is-expr}_{\epsilon \tarrow o}^{o} \, x_\epsilon \Implies
\sembrack{x_\epsilon}_o \Or \Neg \sembrack{x_\epsilon}_o.\] An
instance of LEM may be written in {\HLQE} as
\begin{lstlisting}
`!x:epsilon. isExprType (x:epsilon) (TyBase "bool")
   ==> ((eval x to bool) \/ ~(eval x to bool))`
\end{lstlisting}
that is readily proved.  Instances of this are obtained by applying
\texttt{INST} followed by \texttt{BETA\_REVAL}, the second
beta-reduction rule for evaluations.

\subsection{Induction Schema}

The (first-order) \emph{induction schema for Peano arithmetic} is
usually expressed as the formula schema
\[(P(0) \And \ForallApp x \mdot (P(x) \Implies P(S(x)))) \Implies 
\ForallApp x \mdot P(x)\] where $P(x)$ is a parameterized syntactic
variable that ranges over all formulas of first-order Peano
arithmetic.  If we assume that the domain of the type $\iota$ is the
natural numbers and $\sC$ includes the usual constants of natural
number arithmetic (including a constant $S_{\iota \tarrow
  \iota}$ representing the successor function), then this schema can
be formalized in {\churchqe} as
\begin{align*}
&
\ForallApp f_\epsilon \mdot 
((\mname{is-expr}_{\epsilon \tarrow o}^{\iota \tarrow o} \, f_\epsilon \And
\mname{is-peano}_{\epsilon \tarrow o} \, f_\epsilon) \Implies {} \\
&
\hspace*{2ex}
((\sembrack{f_\epsilon}_{\iota \tarrow o} \, 0 \And
(\ForallApp x_\iota \mdot \sembrack{f_\epsilon}_{\iota \tarrow o} \, x_\iota \Implies
\sembrack{f_\epsilon}_{\iota \tarrow o} \, 
(S_{\iota \tarrow \iota} \, x_\iota)))
\Implies 
\ForallApp x_\iota \mdot \sembrack{f_\epsilon}_{\iota \tarrow o} \, x_\iota))
\end{align*}
where $\mname{is-peano}_{\epsilon \tarrow o} \, f_\epsilon$ holds iff
$f_\epsilon$ represents the syntax tree of a predicate of
first-order Peano arithmetic.  The \emph{induction schema for
  Presburger arithmetic} is exactly the same as the induction schema
for Peano arithmetic except that the predicate
$\mname{is-peano}_{\epsilon \tarrow o}$ is replaced by an appropriate
predicate $\mname{is-presburger}_{\epsilon \tarrow o}$.

It should be noted that the induction schemas for Peano and Presburger
arithmetic are weaker that the full induction principle for the
natural numbers:
\begin{align*}
&
\ForallApp p_{\iota \tarrow o} \mdot 
((p_{\iota \tarrow o} \, 0 \And
(\ForallApp x_\iota \mdot p_{\iota \tarrow o} \, x_\iota \Implies
p_{\iota \tarrow o} \, (S_{\iota \tarrow \iota} \, x_\iota)))
\Implies 
\ForallApp x_\iota \mdot p_{\iota \tarrow o} \, x_\iota)
\end{align*}

\noindent
The full induction principle states that induction holds for all
properties of the natural numbers (which is an uncountable set), while
the induction schemas for Peano and Presburger arithmetic hold only
for properties that are definable in Peano and Presburger arithmetic
(which are countable sets).

The full induction principle is expressed in {\HL} as the
theorem \[\texttt{`!P. P(\_0) /\ (!n. P(n) ==> P(SUC n)) ==> !n. P
  n`}\] named \texttt{num\_INDUCTION}.  However, it is not possible to
directly express the Peano and Presburger induction schemas in {\HL}
without adding new rules of inference to its kernel.

\bsp
The induction schema for Peano arithmetic can be written in
{\HLQE} just as easily as in {\churchqe}:
\begin{lstlisting}
`!f:epsilon. 
   (isExprType (f:epsilon) (TyBiCons "fun" (TyVar "num") (TyBase "bool"))) 
   /\ (isPeano f) 
   ==> 
   (eval (f:epsilon) to (num->bool)) 0 
    /\ (!n:num. (eval (f:epsilon) to (num->bool)) n 
      ==> (eval (f:epsilon) to (num->bool)) (SUC n)) 
   ==> (!n:num. (eval (f:epsilon) to (num->bool)) n)`
\end{lstlisting}
\esp

\noindent \texttt{peanoInduction} is proved from
\texttt{num\_INDUCTION} in {\HLQE} by:

\bsp
\be

  \item Instantiate \texttt{num\_INDUCTION} with
    \texttt{`P:num->bool`} to obtain \texttt{indinst}.

  \item Prove and install the theorem \texttt{nei\_peano} that says the
    variable \texttt{(n:num)} is not effective in \texttt{(eval
      (f:epsilon) to (num->bool))}.

  \item Logically reduce \texttt{peanoInduction}, then prove the
    result by instantiating \texttt{`P:num->bool`} in \texttt{indinst}
    with \texttt{`eval (f:epsilon) to (num->bool)`} using the
    \texttt{INST} rule,  which requires the
    previously proved theorem \texttt{nei\_peano}.

\ee 
\esp 

\noindent 
The induction schema for Presburger arithmetic is stated and proved in
the same way.  By being able to express the Peano and Presburger induction
schemas, we can properly define the first-order theories of Peano
arithmetic and Presburger arithmetic in {\HLQE}.

\section{Related Work}\label{sec:related-work}

Quotation, evaluation, reflection, reification, issues of
intensionality versus extensionality, metaprogramming and
metareasoning each have extensive literature --- sometimes
in more than one field.  For example, one can find a
vast literature on reflection in logic, programming languages,
and theorem proving. Due to space restrictions, we cannot
do justice to the full breadth of issues. For a full discussion,
please see the related work section in~\cite{Farmer18}.
The surveys of Costantini~\cite{Costantini02},
Harrison~\cite{Harrison95} are excellent. From a
programming perspective, the discussion and extensive 
bibliography of Kavvos' D.Phil. thesis~\cite{Kavvos2017}
are well worth reading.

Focusing just on interactive proof assistants, we find
that Boyer and Moore developed a global
infrastructure~\cite{BoyerMoore81} for
incorporating symbolic algorithms into Nqthm~\cite{BoyerMoore88}.
This approach is also used in
ACL2~\cite{KaufmannMoore97}, the successor to Nqthm;
see~\cite{HuntEtAl05}.  Over the last 30 years, the Nuprl group
has produced a large body of work on metareasoning
and reflection for theorem
proving~\cite{AllenEtAl90,Barzilay05,Constable95,Howe92,KnoblockConstable86,Nogin05,Yu07}
that has been implemented in the Nuprl~\cite{Constable86} and
MetaPRL~\cite{HickeyEtAl03} systems.  Proof by reflection has become a
mainstream technique in the Coq~\cite{Coq8.5} proof assistant with the
development of tactics based on symbolic computations like the Coq
ring tactic~\cite{Boutin97,GregoireMahboubi05} and the formalizations
of the \emph{four color theorem}~\cite{Gonthier08} and the
\emph{Feit-Thompson odd-order theorem}~\cite{GonthierEtAl13}.
See~\cite{Boutin97,BraibantPous11,Chlipala13,gonthier2010introduction,GregoireMahboubi05,JamesHinze09,OostdijkGeuvers02}
for a selection of the work done on using reflection in Coq.
Many other systems also support metareasoning and reflection:
Agda~\cite{Norell09,VanDerWalt12,VanDerWaltSwierstra12},
Idris~\cite{Christiansen:2014,Christiansen:2016,Christiansen:2016:Thesis}
Isabelle/HOL~\cite{ChaiebNipkow08},
Lean~\cite{ebner2017metaprogramming},
Maude~\cite{ClavelMeseguer02}, 
PVS~\cite{VonHenkeEtAl98}, 
reFLect~\cite{GrundyEtAl06,MelhamEtAl13},and
Theorema~\cite{GieseBuchberger07,BuchbergerEtAl06}.

The semantics of the quotation operator $\synbrack{\cdot}$ is based on
the \emph{disquotational theory of quotation}~\cite{Quotation12}.
According to this theory, a quotation of an expression $e$ is an
expression that denotes $e$ itself.  In {\churchqe},
$\synbrack{\textbf{A}_\alpha}$ denotes a value that represents the
syntactic structure of $\textbf{A}_\alpha$.  Polonsky~\cite{Polonsky11} 
presents a set of axioms for quotation operators of this
kind.  Other theories of quotation have been proposed --- see
\cite{Quotation12} for an overview.  For instance, quotation can be viewed as
an operation that constructs literals for syntactic values~\cite{Rabe15}.

It is worth quoting Boyer and Moore~\cite{BoyerMoore81} here:
\begin{quote}
The basic premise of all work on extensible
theorem-provers is that it should be possible to add new proof
techniques to a system without endangering the soundness of the system.
It seems possible to divide current work into two broad camps. In the
first camp are those systems that allow the introduction of arbitrary
new procedures, coded in the implementation language, but require that
each application of such a procedure produce a formal proof of the
correctness of the transformation performed. In the second camp are
those systems that contain a formal notion of what it means for a proof
technique to be sound and require a machine-checked proof of the
soundness of each new proof technique. Once proved, the new proof
technique can be used without further justification. 
\end{quote}
This remains true to this day. The systems in the LCF tradition (Isabelle/HOL,
Coq, \HL) are in the ``first camp'',
while Nqthm, ACL2, Nuprl, MetaPRL, Agda, Idris, Lean, Maude and Theorema, 
as well as our approach broadly fall in the ``second camp''. However,
all systems in the first camp have started to offer some reflection
capabilities on top of their tactic facilities. Below we give some 
additional details for each system, leveraging information from the papers 
already cited above as well as the documentation of each system%
\footnote{And some personal communication with some of system authors.}.

\textrm{SSReflect}~\cite{gonthier2010introduction} (\emph{small scale
reflection}) is a Coq extension that works by locally reflecting the
syntax of particular kinds of objects --- such as decidable predicates
and finite structures.  It is the pervasive use of decidability and
computability which gives \textrm{SSReflect} its power, and at the
same time, its limitations.  An extension to PVS allows reasoning much
in the style of \textrm{SSReflect}.  Isabelle/HOL offers a nonlogical
\texttt{reify} function (aka quotation), while its \texttt{interpret}
function is in the logic; it uses global datatypes to represent {\HOL}
terms.

The approach for the second list of systems also varies quite a bit.
Nqthm, ACL2, Theorema (as well as now \HLQE) have global quotation
and evaluation operators in the logic, as well as careful restrictions
on their use to avoid paradoxes. Idris also has global quotation and
evaluation, and the \emph{totality checker} is used to avoid
paradoxes. MetaPRL has evaluation but no global quotation. Agda
has global quotation and evaluation, but their use are mediated by
a built-in \texttt{TC} (TypeChecking) monad which ensures soundness.
Lean works similarly: all reflection must happen in the \texttt{tactic}
monad, from which one cannot escape. Maude appears to offer a global
quotation operator, but it is unclear if there is a global evaluation
operator; quotations are offered by a built-in module, and those are
extra-logical.

\section{Conclusion}\label{sec:conclusion}

{\churchqe}~\cite{Farmer16,Farmer18} is a version of Church's
type theory with global quotation and evaluation operators that is
intended for defining, applying, proving properties about syntax-based
mathematical algorithms (SBMAs), algorithms that manipulate
expressions in a mathematically meaningful ways.  {\HLQE} is an
implementation of {\churchqe} obtained by modifying
{\HL}~\cite{Harrison09}, a compact implementation of the {\HOL} proof
assistant~\cite{GordonMelham93}.  In this paper, we have presented the
design and implementation of {\HLQE}.  We have discussed the
challenges that needed to be overcome.  And we have given some
examples that test the implementation and show the benefits of having
quotation and evaluation in higher-order logic.

The implementation of {\HLQE} was very straightforward since the
logical issues were worked out in {\churchqe} and {\HL} provides good
support for inductive types.  Pleasingly, and surprisingly, no new
issues arose during the implementation. {\HLQE} works in exactly the
same way as {\HL} except that, in the presence of evaluations, the 
instantiation of free variables may require proving side conditions 
that say (1) a variable is not effective in a term or (2) that a variable
represented by a construction is not free in a term represented by a
construction (see subsections~\ref{subsec:cttqe-proof-system}
and~\ref{subsec:hlqe-proof-system}).  This is the only significant
cost we see for using {\HLQE} in place of {\HL}.

{\HLQE} provides a built-in global reflection
infrastructure~\cite{Farmer18}.  This infrastructure can be used
to reason about the syntactic structure of terms and, as we have
shown, to express formula schemas as single formulas.  More
importantly, the infrastructure provides the means to define, apply,
and prove properties about SBMAs.  An SBMA can be defined as a
function that manipulates constructions.  The \emph{meaning formula}
that specifies its mathematical meaning can be stated using the
evaluation of constructions.  And the SBMA's meaning formula can be
proved from the SBMA's definition. In other words, the infrastructure
provides a unified framework for formalizing SBMAs in a proof
assistant.

We plan to continue the development of {\HLQE} and to show that it can
be effectively used to develop SBMAs as we have just described.  In
particular, we intend to formalize in {\HLQE} the example on the
symbolic differentiation we formalized in {\churchqe}~\cite{Farmer18}.
This will require defining the algorithm for symbolic differentiation,
writing its meaning formula, and finally proving the meaning formula
from the algorithm's definition and properties about derivatives.  We
also intend, down the road, to formalize in {\HLQE} the graph of
biform theories encoding natural number of arithmetic described
in~\cite{CaretteFarmer17}.

\bibliography{imps}
\bibliographystyle{splncs04}

%%\setcounter{tocdepth}{1}
%%\listoftodos
%%\setcounter{tocdepth}{0}

\end{document}